\documentclass[a4paper, amsfonts, amssymb, amsmath, reprint, showkeys, nofootinbib, twoside]{revtex4-1}
\usepackage[english]{babel}

\usepackage{array}
\usepackage[utf8]{inputenc}
\usepackage[colorinlistoftodos, color=green!40, prependcaption]{todonotes}
\usepackage{amsthm}
\usepackage{mathtools}
\usepackage{physics}
\usepackage{xcolor}
\usepackage{graphicx}
\usepackage[left=23mm,right=13mm,top=35mm,columnsep=15pt]{geometry} 
\usepackage{adjustbox}
\usepackage{placeins}
\usepackage[T1]{fontenc}
\usepackage{lipsum}
\usepackage{csquotes}

\usepackage{float}
\restylefloat{table}

\usepackage[pdftex, pdftitle={Article}, pdfauthor={Author}]{hyperref} 
\begin{document}
\title{
Accurate calculation 
of the interaction
of a 
barium 
monofluoride 
molecule with an 
argon atom:
A step towards
using matrix 
isolation of 
BaF
for determining 
the electron
electric dipole
moment}

\author{
G. K. Koyanagi, 
R. L. Lambo, 
A. Ragyanszki, 
R. Fournier, 
M. Horbatsch, 
E. A. Hessels}
\email[Correspondence email address: ]{hessels@yorku.ca}
\affiliation{
York University,
Toronto ON,
Canada}
\collaboration{
EDM$^3$
collaboration}
\date{\today} %

\begin{abstract}
Calculations
of the 
BaF-Ar
triatomic
system are 
performed 
with a relativistic 
Hamiltonian
and coupled cluster theory at the
CCSD(T) level
for 
$1386$ 
positions
of the
Ar
atom
relative to the
BaF 
molecule.
Calculations 
are repeated with
increasing basis
sets 
(double-, 
triple-,
quadruple-
and
quintuple-zeta),
and these are 
extrapolated to 
estimate the 
complete-basis-set
limit.
The resulting energies
provide a potential 
energy for the interaction
of an 
Ar
atom 
with a 
BaF
molecule. 
A fit is 
presented that 
parametrizes 
this potential.
This work
is needed for 
an
understanding of
the position,
modes of motion
and 
energy shifts 
of
BaF 
isolated in 
an 
Ar
matrix.
This understanding will
guide the
EDM$^3$
collaboration
in its pursuit of
a precision 
measurement of
the electron
electric dipole
moment
using  
BaF
isolated in 
a cryogenic
Ar
matrix.

\end{abstract}

\keywords{
barium 
monofluoride,
argon, 
triatomic,
matrix isolation,
electron electric dipole moment}

\maketitle

\section{
Introduction}

Barium 
monofluoride
embedded in a 
solid 
argon matrix
is being used
\cite{vutha2018orientation}
by the 
EDM$^3$
collaboration in its pursuit of
a measurement
of unprecedented 
accuracy of the 
electron electric 
dipole moment
(eEDM).
Current limits
\cite{acme2018improved}
on the 
eEDM
already test physics at 
energy scales of up to
100~TeV
and 
constrain 
possible
beyond-the-standard-model
physics that would lead to 
the level of
time-reversal (T) violation 
required to understand the 
asymmetry between matter and
antimatter in the universe.
Measurements of the 
eEDM 
at higher levels of accuracy
will test physics at even
higher energy scales,
and thus may give the first 
indications of the source
of 
T 
violation.

Embedding 
BaF
in an 
Ar
solid 
will
fix the orientation
of the molecules
and
allow for 
large numbers
of 
BaF
to be studied.
To understand 
the effect of the 
matrix on 
the
BaF
molecule, 
the potential energy
between 
this molecule and 
an
Ar 
atom must be 
calculated.
In this work,
we obtain the
ground-state
energies of the 
BaF-Ar 
triatomic
system using 
all-electron
scalar-relativistic
quantum-mechanical
calculations
at the level of
coupled-cluster
theory in the 
Douglas-Kroll-Hess
method.

Recently,
BaF
molecules
have been embedded in both
Ne
\cite{li2022baf}
and
Ar
\cite{corriveau2022baf}
cryogenic solids
by the 
EDM$^3$
collaboration.
In both cases,
the 
spectral features seen with
laser-induced
fluorescence spectroscopy
show indications of the 
internal modes of motion 
of the 
BaF
molecule.
The current work
is needed in order to understand
such modes of motion, 
as well as for
understanding of the 
the position of the 
BaF within the 
FCC 
Ar
crystal
and of the
energy shifts 
of
the
BaF
energy levels
by the 
Ar
matrix.
This understanding will
guide the
EDM$^3$
collaboration
in pursuing 
a precision 
measurement of
the electron
electric dipole
moment
using  
BaF
isolated in 
a cryogenic
Ar
matrix.

\section{
Methods}
\begin{figure}
\includegraphics
[width=0.9\linewidth]{
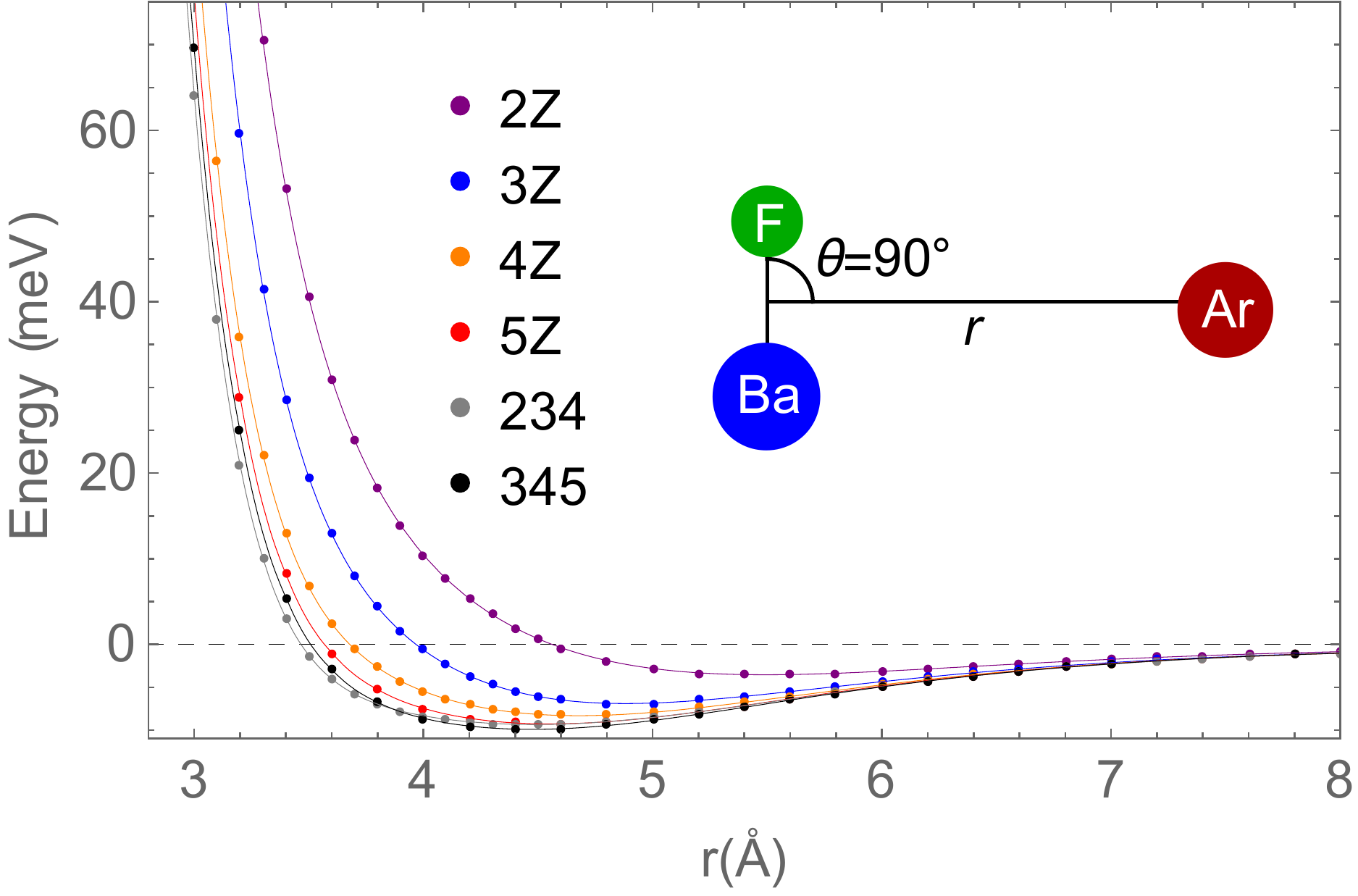}
\caption{
(color
online)
The calculated binding
energy of the 
BaF-Ar
system
for a series of distances
$r$
for the case of 
$\theta=90^\circ$.
The 
$\zeta=2$
through
$5$
calculations are shown,
along with the 
extrapolations of
Eqs.~(\ref{eq:234})
and 
(\ref{eq:345}).
The close agreement between
the two extrapolations indicates
the level of accuracy of the 
calculations.
}
\label{fig:extrap90}
\end{figure}

Calculations of 
the 
BaF-Ar
binding energy 
are
performed at 
1386
positions 
of the 
Ar 
relative to the 
BaF
(22 
angles,
$\theta$,
and 
63 
distances,
$2.6$~\AA~$\le r \le16$~\AA,
where 
$r$
is  
the distance between
the 
Ar 
nucleus
and
the 
midpoint between
the 
Ba
and
F
nuclei
and 
$\theta$
is defined
to be the angle
between the
Ar 
and 
F
nuclei
relative to this 
midpoint, 
as shown in 
the inset in
Fig.~\ref{fig:extrap90}).
As described below,
these 
quantum-mechanical 
calculations
are 
scalar-relativistic,
all-electron,
include correlation at the 
CCSD(T) 
level of theory,
and are 
extrapolated to 
a complete basis
set.

For 
F
and 
Ar,
non-relativistic 
all-electron 
$\zeta=2$
through
$5$
basis sets 
(aug-cc-pV$\zeta$Z) 
are obtained
from 
Kendall et al.
\cite{kendall1992electron}
and
Woon et al. 
\cite{woon1993gaussian}.
These
basis sets were 
fully 
uncontracted 
and atomic calculations
were performed at the 
Douglas-Kroll-Hess 
(DKH) 
second-order
scalar-relativistic level. 
Using
orbital occupations
from this 
calculation, 
the basis sets were 
re-contracted 
to their
original structure. 
The contraction coefficients 
for these 
basis sets
varied by 
about 
2\%
from the original 
non-relativistic 
basis set 
for functions that make 
their largest contributions
near the 
radius of highest 
electron probability.
For 
Ba
$\zeta=2$
through 4,
aug-cc-pV$\zeta$Z-DK3
all-electron 
relativistic 
basis sets 
were taken from 
Hill and Peterson \cite{hill2017gaussian}
These 
third-order 
Douglas-Kroll-Hess 
basis sets
were used 
under
the 
second-order
DKH
method,
as it was
found to have
negligible differences
in its 
contraction coefficients. 
A new 
quintuple-zeta
basis set for 
Ba
(aug-cc-pV5Z-DKH)
was constructed 
for this study.
This basis set 
was developed from
the 
$\zeta=4$ 
basis set 
by augmenting the 
outer-shell
flexibility by adding 
uncontracted 
s-,
p-, 
and
d-basis
functions 
at the next smaller
radius than those 
already in the 
uncontracted 
$\zeta=4$
basis set. 
Further, 
the 
$\zeta=4$
polarization
and 
augmentation functions 
were replaced with the 
$\zeta=5$
polarization 
and 
augmentation functions 
from Hill and Peterson
\cite{hill2017gaussian}.
The contraction scheme
for the resulting 
aug-cc-pV5Z-DKH 
basis set can be denoted 
[36s,~31p,~21d,~7f,~3g,~2h]
$\to$
[12s,~10p,~8d,~5f,~3g,~2h] 
(i.e.,
the 
36
s-type
primitive functions
were contracted into 
12
linear combinations,
etc.).
This basis set is given 
in 
Tables~\ref{table:S5Z}
through 
\ref{table:H5Z}
of the Supplementary materials.

For the 
triatomic 
calculations, 
the binding energies 
calculated are
$E($BaF-Ar$)-E($BaF$)-E($Ar$)$.
The use of 
a finite basis set results
in 
$E($BaF-Ar$)$
being artificially low 
due to diffuse
Ar
orbitals improving 
the core description
of
BaF
and diffuse 
Ba
and 
F 
orbitals improving 
the core description of 
Ar.
To correct for this
the standard 
counterpoise correction
method was used, 
whereby 
$E($Ar$)$ 
is computed with
the inclusion of 
BaF 
ghost orbitals 
and 
$E($BaF$)$
is computed with 
the inclusion of 
Ar 
ghost
orbitals
\cite{boys1970calculation}.

The correlation energy
is incorporated using 
the 
coupled-cluster 
method 
CCSD(T) 
\cite{raghavachari1989fifth,bartlett1990non,stanton1997ccsd}.
The 
correlation-consistent
basis sets are
designed to 
recover 
the
correlation energy
in a consistent
manner across 
atom 
centers
and in a
progressive manner 
as valence and
polarization
orbitals are added. 
As such, 
extrapolation
to total correlation
energy recovery
can be achieved
by studying the
problem with a
succession of
basis sets. 
Extrapolation to a
complete-basis-set 
(CBS)
of energies
$E_2$,
$E_3$
and
$E_4$
(from
$\zeta=2$,
$3$
and
$4$
basis sets)
can be obtained
following the work 
of 
Buchachenko
and
Viehland
\cite{buchachenko2018interaction}:
\begin{eqnarray}
E_{234}
\!=\! 
1.677
E_4
\!-\! 
0.712
E_3
\!+\! 
0.035
E_2
\label{eq:234}
\end{eqnarray}
Similarly,
one can extrapolate
from
$\zeta=3$,
$4$
and
$5$
basis sets:
\begin{eqnarray}
E_{345}
\!=\!
1.593 E_5
\!-\!
0.597 E_4 
\!+\!
0.004 E_3.
\label{eq:345}
\end{eqnarray}
Figure~\ref{fig:extrap90} 
shows 
$\zeta=2$
through
$5$
calculations
for the 
BaF-Ar 
system
for the case of 
$\theta=90^\circ$,
along with the two
extrapolations.
Note the good  
agreement between the
two extrapolations.
The 
computationally-expensive
$\zeta=5$
calculations were performed
at 
361
carefully-chosen 
points
and,
at points where 
the 
$\zeta=5$
calculation
were not performed,
$E_{345}$
can be 
estimated to high accuracy
from a smooth interpolation
of 
$E_{345}-E_{234}$
obtained from the
361 points.
We take 
$E_{345}$
as our best 
estimate of the 
potential and 
we make an
estimate of the 
uncertainty based 
on one quarter of 
the difference 
between
$E_{234}$
and
$E_{345}$.

The current 
calculations are 
performed with the 
internuclear 
Ba-F
distance
fixed at
2.16~\AA
--
the separation 
determined from 
rotational spectroscopy
\cite{bernard1992laser}.
The 
BaF
binding energy 
(6 eV
\cite{ehlert1964mass,hildenbrand1968mass})
is large compared to the 
BaF-Ar
binding energy 
(23 meV; 
calculated here),
which leads to a much 
stronger restoring 
force for 
BaF 
stretching
compared to 
BaF-Ar
interactions
and
justifies
fixing 
the
BaF
internuclear 
separation.

Our 
calculational 
methods
were tested 
by
using the same
techniques and basis sets
to calculate
properties of 
the 
BaF 
molecule,
of the 
Ar 
and 
Ba 
atoms,
and 
of the 
Ba$^+$
ion,
as summarized
in 
Table~\ref{table:Compare}.
Our calculations yield
a 
BaF
dissociation energy 
$D_{\rm e}=$~5.95(3)~eV,
where the uncertainty 
is one quarter of the 
$E_{345}-E_{234}$
difference.
This is in good agreement with
the measured values of 
6.05(7)~eV
\cite{hildenbrand1968mass}
and 
5.85(9)~eV
\cite{ehlert1964mass,hildenbrand1968mass}.
Our calculated
BaF
equilibrium distance
is
$R_{\rm e}=$~2.172(2)~\AA,
which agrees at the 
0.5\%
level with
the experimental value of
2.1592~\AA,
from 
$B=$0.21652967(7)~cm$^{-1}$
\cite{bernard1992laser,ryzlewicz1980formation,le2011determining}.
Our calculated 
vibrational 
constant
$\omega_{\rm e}=$~467(2)~cm$^{-1}$
(determined from fits of the 
calculated points near the potential 
minimum to a Morse potential)
is
in good agreement with the 
experimental value of 
469.4161(19)~cm$^{-1}$
\cite{bernard1992laser}.
Finally,
our calculated value for the 
permanent dipole moment of 
BaF
is 
3.25(7)~debye,
which
agrees with the 
measured value
\cite{ernst1986hyperfine}
of
3.170(3)~debye.

\begin{table}[hbt!]
\begin{ruledtabular}
\caption{\label{table:Compare} 
Comparisons of 
our calculations
with 
experimental values for
the
dissociation energy, 
$D_{\rm e}$,
equilibrium distance,
$R_{\rm e}$,
vibrational 
constant,
$\omega_{\rm e}$,
and
the permanent dipole moment,
$d_{\rm p}$,
of 
BaF,
and for the 
dipole polarizabilities,
$\alpha$,
of
Ar,
Ba
and
Ba$^+$. 
Uncertainties in the 
last digits are shown in parentheses. 
Our calculations show good agreement
with experimental values.
}
\begin{tabular}{lll}
&
Experiment&
This work\\
\hline
$D_{\rm e}$(eV)&
6.05(7)\cite{hildenbrand1968mass},
5.85(9)\cite{ehlert1964mass,hildenbrand1968mass}
&5.95(3)\\
$R_{\rm e}$(\AA)&
2.1592\cite{bernard1992laser,ryzlewicz1980formation,le2011determining}&
2.172(2)\\
$\omega_{\rm e}$(cm$^{-1}$)&
469.4161(19)\cite{bernard1992laser}&
467(2)\\
$d_{\rm p}$(debye)&
3.170(3)\cite{ernst1986hyperfine}&
3.25(7)\\
$\alpha_{\rm Ar}$($a_0^3$)& 
11.083(7)\cite{schwerdtfeger20192018}&
11.05(5)\\
$\alpha_{\rm Ba}$($a_0^3$)& 
268(22)\cite{schwartz1974measurement}& 
268(8)\\
$\alpha_{\rm Ba^+}$($a_0^3$)& 
123.88(5)\cite{snow2007fine}
&
126(4) 
\end{tabular}
\end{ruledtabular}
\end{table}

The 
long-distance 
behaviour of the 
BaF-Ar 
potential is
largely due to the 
dipole 
polarizabilities
of the constituents.
Since 
Ba 
is much more 
polarizable 
than
F,
and because the 
Ba 
atom
in
BaF 
is partially ionized, 
we calculate the 
polarizabilities
of 
Ar, 
Ba
and 
Ba$^+$
using the methods
described in this work
and compare
these calculations
to 
experimental values.
For 
Ar 
we calculate
11.05(5)~$a_0^3$,
which is in good 
agreement with 
the experimental
value of 
11.083(7)~$a_0^3$
\cite{schwerdtfeger20192018}.
For
Ba,
we calculate a
polarizability 
of 
268(8)~$a_0^3$,
which agrees well 
with the experimental 
value of
268(22)~$a_0^3$
\cite{schwartz1974measurement},
and for 
Ba$^+$
we calculate
126(4)~$a_0^3$,
which agrees well with 
the experimental value 
\cite{snow2007fine}
of
123.88(5)~$a_0^3$.
The agreement with experiment for 
the 
BaF 
potential 
and 
the 
polarizabilities
(as
summarized in 
Table~\ref{table:Compare})
justifies our uncertainty 
estimate 
based on 
$\frac{1}{4}(E_{345}-E_{234})$.

\begin{figure}
\includegraphics
[width=0.9\linewidth]{
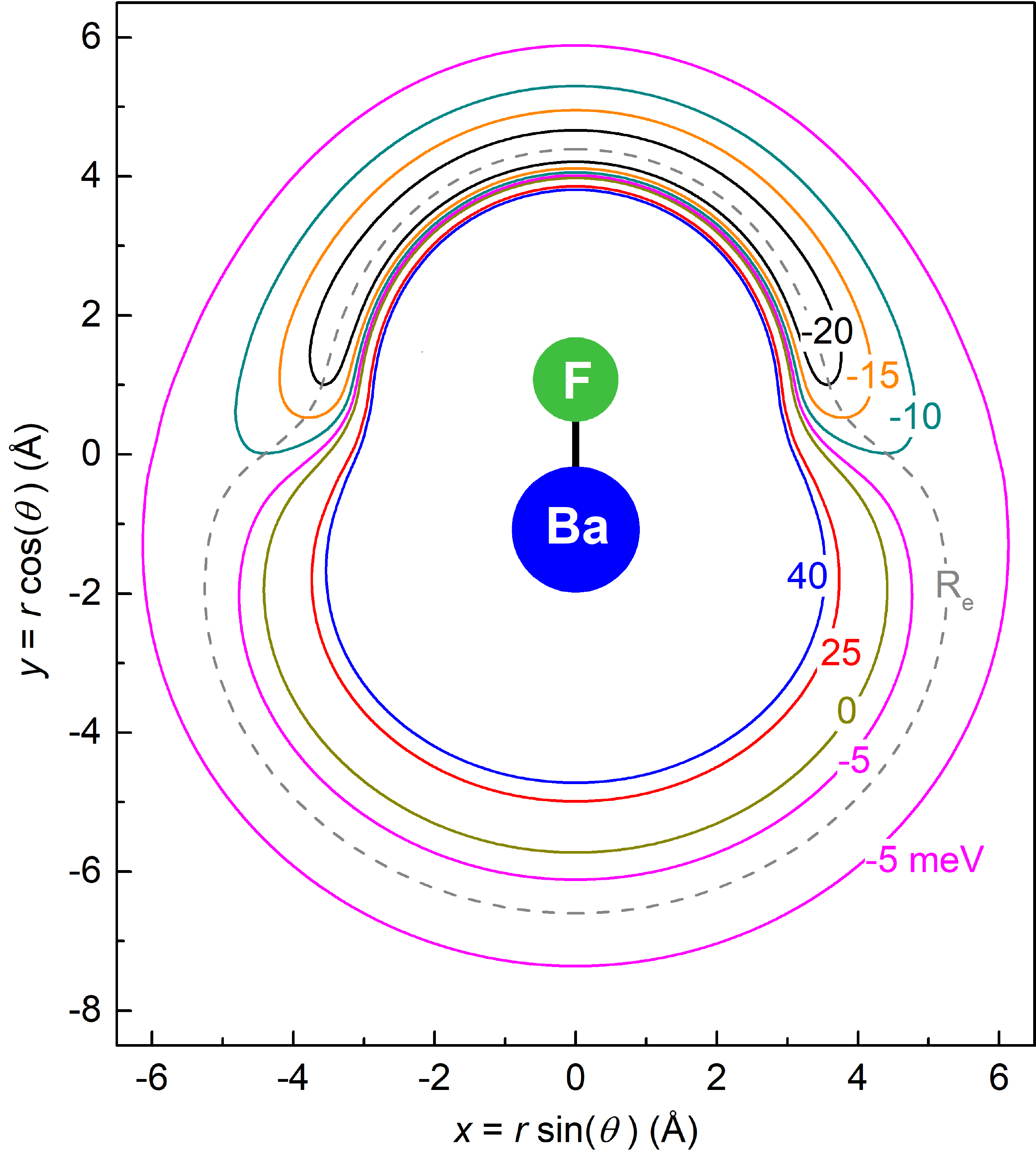}
\caption{
(color online)
A contour graph of 
the calculated 
potential energy between 
an
Ar
atom and a 
BaF 
molecule.
The geometric centre of the 
Ba 
and 
F
nuclei
is situated at 
$(x,y)=(0,0)$.
}
\label{fig:contours}
\end{figure}

\section{
Results}
\begin{figure*}[hbt!]
    \includegraphics[width=1.0\linewidth]{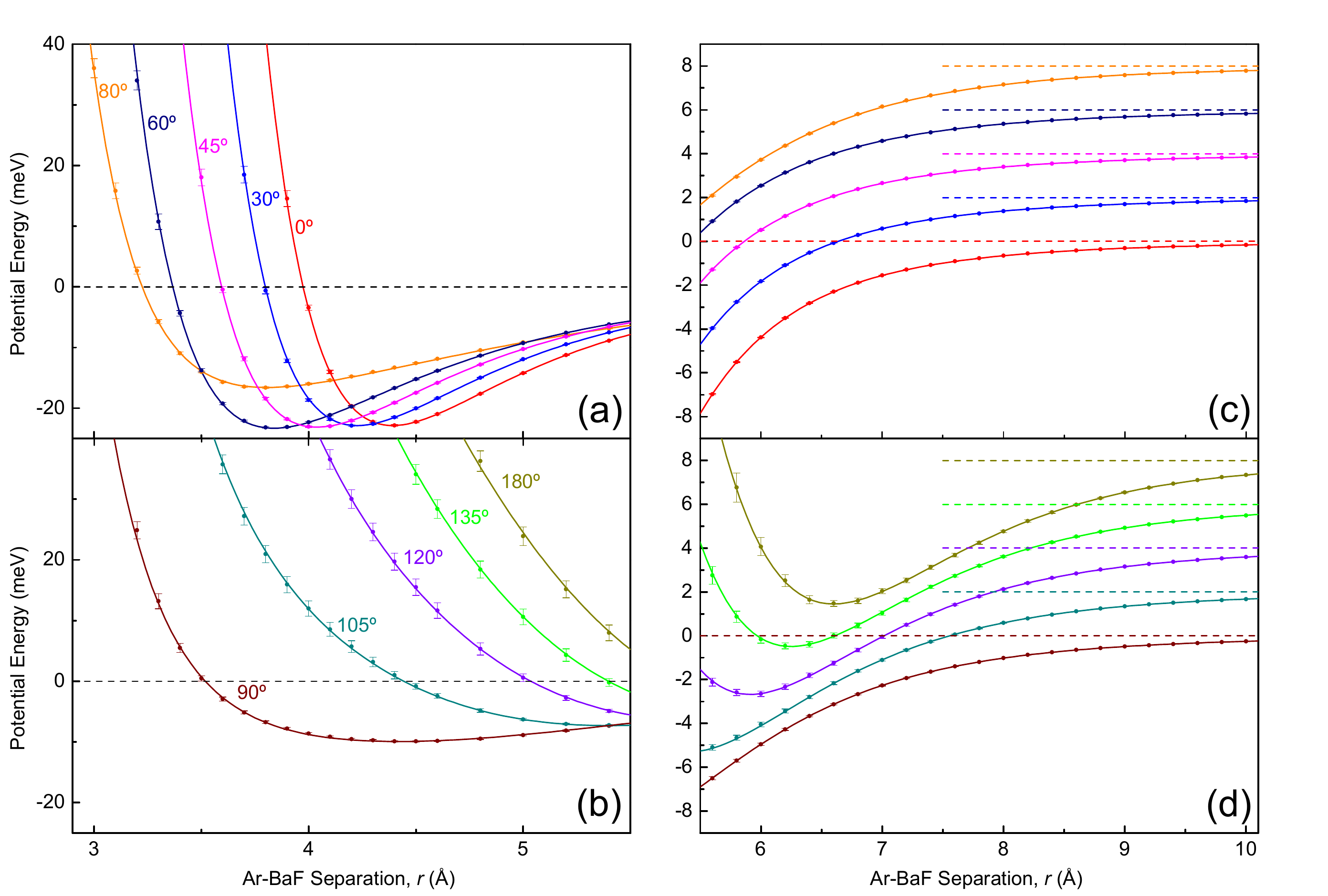}
    \caption{
    (color 
    online)
    The potentials shown for
    10
    of the 
    22
    angles,
    $\theta$,
    calculated here.
    The points plotted 
    are the 
    $E_{345}$
    extrapolations,
    and the uncertainties
    are one quarter of the 
    typical 
    $E_{345}-E_{234}$
    difference at this 
    energy.
    The red fit curves 
    are 
    as described in 
    Section~\ref{sec:fit}.
    The potential curves 
    at longer distances 
    shown in 
    (c)
    and
    (d)
    are offset for clarity.
    }
    \label{fig:potentials}
\end{figure*}

Figure~\ref{fig:contours} 
shows a contour plot of 
our calculated 
BaF-Ar 
potential and
Figure~\ref{fig:potentials}
shows the calculated
BaF-Ar
energies for 
10
of the
22
angles
$\theta$
used.
Parts 
(a)
and
(b)
of 
Fig.~\ref{fig:potentials}
show the 
shorter-distance
potentials
and parts
(c)
and 
(d) 
show the 
longer-distance 
potential, 
and these have 
their zeros offset
by 
2,
4,
6 
and 
8~meV
for clarity.
The uncertainties shown 
in the figure are based
on one quarter of typical 
$E_{345}-E_{234}$
differences,
with uncertainties of 
1.5~meV
at 
25~meV on the repulsive wall,
0.5~meV
at 
$E_{345}=0$,
0.1~meV
at 
the 
potential minimum,
0.04~meV
half way up the attractive 
part of the potential,
and reducing to 
0.005~meV 
at long distances.
The full set of 
calculations is 
given in 
Tables~\ref{table:sup0}
through
\ref{table:sup180}
of the supplementary
materials.
The solid lines
in the figure are 
based on 
the fit described 
in  
Section~\ref{sec:fit}.

The 
Ar-BaF
potential varies 
strongly with angle.
For 
$\theta=0$,
for which the 
Ar
is on the 
F
side of the 
BaF
molecule,
the interaction is 
relatively strong,
with a binding energy
$D_{\rm e}=$
22.8(1)~meV
at 
$R_{\rm e}\approx$4.4~\AA
from 
the geometric centre
of the 
Ba
and 
F
nuclei
(see 
Fig.~\ref{fig:contours}).
The 
Ar 
atom is bound more weakly 
on the 
Ba
side 
and
the equilibrium 
distance is much larger
($D_{\rm e}=$
6.5(1)~meV;
$R_{\rm e}\approx$
6.6~\AA).
The 
repulsion at shorter 
distances shows
a steep curve at
small
$\theta$
and 
an increasingly 
gentle curve for
larger 
$\theta$.
As expected,
the 
larger 
Ba
atom leads to 
the repulsive wall
of the potential energy 
being at larger 
distances 
(with a value of
40~meV 
at
approximately 
4.75~\AA)
and
the smaller
F
atom allows for 
closer separations
(40~meV
at 
approximately
3.8~\AA).
The repulsive wall is 
even closer near 
$\theta=90^\circ$,
where 
40~meV occurs
at approximately
3~\AA.


\section{
Parametrization of the 
$\textbf{{\rm BaF-Ar}}$
potential energy 
\label{sec:fit}}

The 
BaF-Ar
potential energy
can be 
parametrized 
in terms of 
a generalized form
of the 
Morse/long-range
(MLR)
potential
\cite{LeRoyPotential2007,LeRoy3D2008,li2012intramolecular,zhai2018constructing}.
As shown by the red curves in 
Fig.~\ref{fig:potentials},
for each angle
$\theta$,
the potential 
is fit to 
\begin{eqnarray}
\label{eq:MLR}
V(r)
=
\Big[
\Big(
&1&
-
\frac
{U_{\rm LR}(r)}
{U_{\rm LR}
(
R_{\rm e})}
e^
{-\beta(r)
X_5(r,
R_{\rm e}
)}
\Big)^2
-
1
\Big]   
D_{\rm e}.
\end{eqnarray}
Here,
$R_{\rm e}$
and 
$D_{\rm e}$
set the 
equilibrium position 
and 
binding energy,
respectively,
and  
$U_{\rm LR}$
is the 
long-range
potential, 
which can be approximated
as
\begin{eqnarray}
U_{\rm LR}(r) 
&=&
d_6(r)
\frac
{
C_6
}
{r^6}
,
\end{eqnarray}
where 
$d_6(r)$
is a damping function
that
cuts off the 
potential 
at small
$r$.
The damping function
is a
generalized 
version
\cite{leRoy2011long} 
of the 
Tang-Toennies 
damping 
function
\cite{tang1984improved}
\begin{eqnarray}
d_m(r)
&=& 
1
-
e
^
{-r/R_{\rm d}}
\sum_{k=0} ^{m-1}
\frac
{(r/R_{\rm d})^k}
{k!},
\end{eqnarray}
where
$R_{\rm d}$
is the damping distance.

The exponential in 
Eq.~(\ref{eq:MLR})
employs a 
switching function
\begin{eqnarray}
X_n(r,R)
&=&
\frac
{r^n-R^n}
{r^n+R^n},
\end{eqnarray}
to scale 
$r$,
and
\begin{eqnarray}
\label{eq:beta}
\beta(r)
&=&
X_5(r,
R_{\rm ref})
\ln\frac
{2 
D_{\rm e}
}
{U_{\rm LR}
(R_{\rm e})}
\nonumber
\\
&+&
(1
-
X_5(r,
R_{\rm ref}))
\sum_{i=0}^2
\beta_i
X_3(r,
R_{\rm ref})
^i,
\end{eqnarray}
with
$R_{\rm ref}$
set to 
6.60~\AA,
the
value of 
$R_{\rm e}$
at 
$\theta=180^\circ$
(the angle at 
which 
$R_{\rm e}$
has its 
maximum value).

\begin{figure}[hbt!]
\includegraphics[width=0.95\linewidth]{
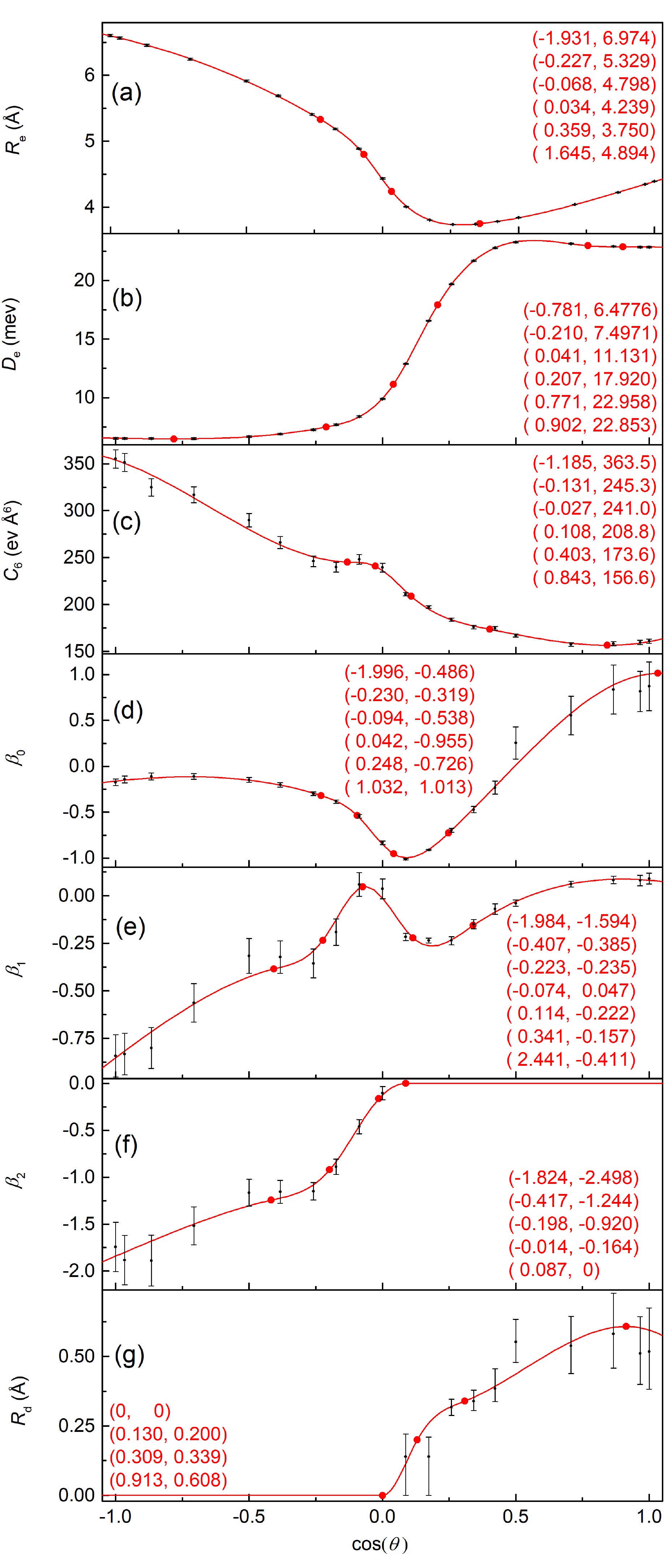}
\caption{
(color online)
The seven parameters
obtained from
fits of 
$E_{345}$
BaF-Ar 
binding energies
to
the
MLR
potential
are shown
as 
data points 
with error bars 
for the 
22
angles
$\theta$
used in this work.
The parameters 
vary smoothly
with
$\theta$
and the red curves
show the 
best-fit
cubic splines
that are used to fit
the calculated
potential at 
all distances 
$r$
and 
all angles
$\theta$.
The coordinates 
of the 
nodes
for these splines are
shown in red.
}
\label{fig:sevenPar}
\end{figure}

The calculated potentials
can be fit to this
MLR
potential for each 
angle 
$\theta$
to determine
the values of the 
seven parameters:
$R_{\rm e}$,
$D_{\rm e}$,
$C_6$,
$R_{\rm d}$,
$\beta_0$,
$\beta_1$
and
$\beta_2$
for that
particular value of
$\theta$.
Here, 
$r$
and 
$\theta$
are defined 
relative to the 
center 
point
between the 
Ba
and 
F
nuclei,
as shown in 
the inset in 
Fig.~\ref{fig:extrap90}.
The seven fit 
parameters vary 
smoothly with 
$\theta$, 
as can be seen in 
Fig.~\ref{fig:sevenPar}.
As the fits return values
of very nearly zero for 
$R_{\rm d}$
for 
$\theta \le 90^\circ$
and 
for
$\beta_2$
for 
$\theta>90^\circ$,
and since these
parameters are highly 
correlated, we set 
these parameters to zero
in these ranges, 
as shown in 
Fig.~\ref{fig:sevenPar}.
This change of
parametrization 
at 
$\theta=90^\circ$
results from the 
shape of contour lines 
in 
Fig.~\ref{fig:contours}
abruptly changing near 
$(x=4, y=0)$.

One approach to describing
the smooth variation 
of the seven parameters
would
be to use an expansion
in terms of 
Legendre polynomials
(as, 
for example,
in 
Ref.~\cite{sayfutyarova2012interactions}).
However, 
the flat regions, 
particularly for 
$D_{\rm e}$,
$R_{\rm d}$,
$\beta_1$
and
$\beta_2$
in 
Fig.~\ref{fig:sevenPar},
make such an expansion 
nonideal, 
since the expansion
adds nonphysical oscillations
in the flat regions
and a large number 
of 
Legendre 
polynomials would 
have to be included
to reduce these
oscillations.

A better approach
is to perform a 
least-squares 
fit 
of the calculated 
potential energies to 
$n$-knot
cubic splines
for each of the parameters.
Figure~\ref{fig:sevenPar}
shows 
(in red)
such splines.
Also shown 
in the figure
are 
the 
$n=4$ 
to
$7$
knots that 
define these curves.
The curves can
be obtained from the knots
using the standard
definition of the 
cubic spline
(see, 
for example,
Ref.~\cite{press2007numerical}).
These knots therefore 
(along with 
Eqs.~(\ref{eq:MLR})-(\ref{eq:beta})
define 
a continuous function that gives
the
full potential 
at all 
$r$ 
and
$\theta$.
This analytic form 
is useful for summarizing
the results of our calculation
and 
could be useful when 
these results are used for
modelling the
local environment of a 
BaF
molecule
in an 
Ar
solid.

\section{
conclusions}

Accurate 
ground-state
BaF-Ar 
interaction energies 
and their uncertainties
have been
calculated for a range of 
angles
and
distances.
The calculated energies 
are fit to 
a functional 
form  
(Eqs.~(\ref{eq:MLR})-(\ref{eq:beta}),
with spline interpolation 
of parameters as a 
function of angle)
which 
may 
also 
be useful for other 
diatomic molecules 
interacting with a 
rare-gas
atom.
These calculations are 
critical to understanding the 
local environment of a 
ground-state
BaF
molecule that
is embedded into
a cryogenic
Ar 
solid,
and may also be 
useful for 
understanding 
gas-phase
Ar-BaF
collisions.

Future work will extend this 
understanding to excited 
states of 
BaF,
allowing for an 
understanding of the 
matrix-induced
shifts of optical
spectrum and of 
the modes of motion
of the excited state.
This work is a necessary 
step towards the 
EDM$^3$
program for 
measuring the electron 
electric dipole moment
using 
embedded 
BaF
molecules.

\section{acknowledgements}
This work is supported by
the 
Gordon and Betty Moore Foundation,
the
Alfred P. Sloan Foundation,
the 
Templeton Foundation
in conjunction with the 
Northwestern Center for Fundamental Physics,
the
Natural Sciences and Engineering Research Council
of Canada
and
York University.
Computations for this work were 
enabled by support provided by 
Compute Canada.

\clearpage

\bibliography{BaFAr.bib}

\newpage
\section{Supplementary Materials}
Tables~\ref{table:sup0}
to
\ref{table:sup180}
show the 
calculated 
potential energies 
$V(r,\theta)$ 
(in meV) 
for the basis sets 
$\zeta=$
2 through 5
for angles 
$\theta$
given in the 
table captions.
Also shown 
are the 
complete-basis-set
extrapolations of 
Eqs.~(\ref{eq:234}) 
and 
(\ref{eq:345}).

Values in italics
are 
interpolations 
or extrapolations.
The italics values 
for 
$r<10$~\AA~are 
based on the 
$E_{234}$ 
values
plus 
an interpolation 
of 
$E_{345}-E_{234}$.
For
$r>10$~\AA,
the values in italics
are 
a smooth
extrapolation 
of  
intermediate-range
data
using a fit
to inverse powers
of 
$r$.

Tables~\ref{table:S5Z}
through
\ref{table:H5Z}
give the contraction 
coefficients
for our
$\zeta=5$
Ba 
basis set.

\clearpage

\begin{table} \begin{ruledtabular}
\setlength\extrarowheight{-2pt}
\scriptsize
\caption{Potential energy $E(r,\theta)$ in meV for the basis sets  $\zeta=$ 2 through 5 and for the  complete-basis-set extrapolations of  Eqs.~(\ref{eq:234})  and  (\ref{eq:345}) for $\theta=$ 0$^\circ$.}
\label{table:sup0}
\begin{tabular}
{rrrrrrrr}
$r$(\AA)&
$E(2\zeta)$&
$E(3\zeta)$&
$E(4\zeta)$&
$E(5\zeta)$&
$E(234)$&
$E(345)$\\ \hline
2.6&10\,259.9&9\,589.5&9\,376.9&---&9\,256.5&\textit{9\,220.3}\\
2.7&7\,365.1&6\,882.6&6\,724.5&---&6\,634.3&\textit{6\,605.0}\\
2.8&5\,212.5&4\,858.2&4\,738.5&---&4\,669.8&\textit{4\,646.5}\\
2.9&3\,651.1&3\,387.1&3\,294.7&---&3\,241.4&\textit{3\,223.1}\\
3.0&2\,534.2&2\,336.5&2\,264.5&---&2\,222.6&\textit{2\,208.6}\\
3.1&1\,743.3&1\,595.2&1\,538.9&---&1\,505.9&\textit{1\,495.6}\\
3.2&1\,188.0&1\,076.9&1\,032.9&---&1\,007.0&\textit{999.6}\\
\end{tabular}
\end{ruledtabular} \end{table} 
\begin{table} \begin{ruledtabular}
\scriptsize \caption{Potential energy $E(r,\theta)$ in meV for the basis sets  $\zeta=$ 2 through 5 and for the  complete-basis-set extrapolations of  Eqs.~(\ref{eq:234})  and  (\ref{eq:345}) for $\theta=$ 5$^\circ$.}
\begin{tabular}
{rrrrrrrr}
$r$(\AA)&
$E(2\zeta)$&
$E(3\zeta)$&
$E(4\zeta)$&
$E(5\zeta)$&
$E(234)$&
$E(345)$\\ \hline
2.6&10\,033.0&9\,378.7&9\,170.2&---&9\,051.9&\textit{9\,016.8}\\
2.7&7\,198.1&6\,726.5&6\,571.2&---&6\,482.5&\textit{6\,454.0}\\
2.8&5\,093.3&4\,746.5&4\,628.7&---&4\,561.0&\textit{4\,538.2}\\
2.9&3\,567.3&3\,308.7&3\,217.6&---&3\,165.0&\textit{3\,147.2}\\
3.0&2\,475.9&2\,282.1&2\,211.1&---&2\,169.8&\textit{2\,156.1}\\
3.1&1\,703.1&1\,557.8&1\,502.3&---&1\,469.7&\textit{1\,459.6}\\
3.2&1\,160.5&1\,051.4&1\,008.0&987.3&982.4&975.2\\
\end{tabular}
\end{ruledtabular} \end{table} 
\begin{table} \begin{ruledtabular}
\scriptsize \caption{Potential energy $E(r,\theta)$ in meV for the basis sets  $\zeta=$ 2 through 5 and for the  complete-basis-set extrapolations of  Eqs.~(\ref{eq:234})  and  (\ref{eq:345}) for $\theta=$ 15$^\circ$.}
\begin{tabular}
{rrrrrrrr}
$r$(\AA)&
$E(2\zeta)$&
$E(3\zeta)$&
$E(4\zeta)$&
$E(5\zeta)$&
$E(234)$&
$E(345)$\\ \hline
2.6&8\,402.7&7\,861.3&7\,681.9&---&7\,579.4&\textit{7\,552.0}\\
2.7&6\,008.6&5\,613.1&5\,477.5&---&5\,399.5&\textit{5\,377.2}\\
2.8&4\,246.6&3\,952.6&3\,848.2&---&3\,787.8&\textit{3\,770.0}\\
2.9&2\,972.6&2\,752.0&2\,670.7&---&2\,623.3&\textit{2\,609.3}\\
3.0&2\,062.1&1\,896.1&1\,832.5&---&1\,795.2&\textit{1\,784.5}\\
3.1&1\,417.3&1\,292.4&1\,242.6&---&1\,213.2&\textit{1\,205.2}\\
3.2&964.7&870.2&831.3&813.1&808.3&802.5\\
\end{tabular}
\end{ruledtabular} \end{table} 
\begin{table} \begin{ruledtabular}
\scriptsize \caption{Potential energy $E(r,\theta)$ in meV for the basis sets  $\zeta=$ 2 through 5 and for the  complete-basis-set extrapolations of  Eqs.~(\ref{eq:234})  and  (\ref{eq:345}) for $\theta=$ 30$^\circ$.}
\begin{tabular}
{rrrrrrrr}
$r$(\AA)&
$E(2\zeta)$&
$E(3\zeta)$&
$E(4\zeta)$&
$E(5\zeta)$&
$E(234)$&
$E(345)$\\ \hline
2.6&4\,766.8&4\,453.3&4\,336.7&---&4\,268.8&\textit{4\,251.5}\\
2.7&3\,394.2&3\,157.2&3\,065.8&---&3\,012.3&\textit{2\,998.4}\\
2.8&2\,394.0&2\,213.6&2\,141.8&---&2\,099.5&\textit{2\,088.4}\\
2.9&1\,672.9&1\,535.2&1\,478.8&---&1\,445.5&\textit{1\,436.8}\\
3.0&1\,158.0&1\,052.4&1\,008.2&---&981.9&\textit{975.2}\\
3.1&793.4&711.9&677.2&---&656.6&\textit{651.6}\\
3.2&537.4&473.7&446.5&434.0&430.4&426.7\\
\end{tabular}
\end{ruledtabular} \end{table} 
\begin{table} \begin{ruledtabular}
\scriptsize \caption{Potential energy $E(r,\theta)$ in meV for the basis sets  $\zeta=$ 2 through 5 and for the  complete-basis-set extrapolations of  Eqs.~(\ref{eq:234})  and  (\ref{eq:345}) for $\theta=$ 45$^\circ$.}
\begin{tabular}
{rrrrrrrr}
$r$(\AA)&
$E(2\zeta)$&
$E(3\zeta)$&
$E(4\zeta)$&
$E(5\zeta)$&
$E(234)$&
$E(345)$\\ \hline
2.6&2\,086.5&1\,924.3&1\,855.7&---&1\,814.9&\textit{1\,804.4}\\
2.7&1\,481.5&1\,355.0&1\,300.7&---&1\,268.3&\textit{1\,259.8}\\
2.8&1\,042.1&942.7&899.8&---&874.2&\textit{867.4}\\
2.9&725.8&646.9&613.0&---&592.9&\textit{587.6}\\
3.0&500.0&436.7&409.9&397.3&394.0&389.9\\
3.1&339.9&288.6&267.4&257.5&254.8&251.7\\
3.2&227.3&185.2&168.4&160.7&158.5&156.1\\
\end{tabular}
\end{ruledtabular} \end{table} 
\begin{table} \begin{ruledtabular}
\scriptsize \caption{Potential energy $E(r,\theta)$ in meV for the basis sets  $\zeta=$ 2 through 5 and for the  complete-basis-set extrapolations of  Eqs.~(\ref{eq:234})  and  (\ref{eq:345}) for $\theta=$ 60$^\circ$.}
\begin{tabular}
{rrrrrrrr}
$r$(\AA)&
$E(2\zeta)$&
$E(3\zeta)$&
$E(4\zeta)$&
$E(5\zeta)$&
$E(234)$&
$E(345)$\\ \hline
2.6&838.7&741.2&694.2&---&665.9&\textit{661.4}\\
2.7&593.4&514.2&476.9&---&454.5&\textit{451.0}\\
2.8&416.0&350.9&321.2&308.3&303.4&300.8\\
2.9&288.4&234.5&210.8&200.5&196.6&194.6\\
3.0&197.2&152.2&133.1&125.0&121.8&120.3\\
3.1&132.38&94.62&79.24&72.73&70.15&68.93\\
3.2&86.60&54.81&42.34&37.08&35.01&34.02\\
\end{tabular}
\end{ruledtabular} \end{table} 
\begin{table} \begin{ruledtabular}
\scriptsize \caption{Potential energy $E(r,\theta)$ in meV for the basis sets  $\zeta=$ 2 through 5 and for the  complete-basis-set extrapolations of  Eqs.~(\ref{eq:234})  and  (\ref{eq:345}) for $\theta=$ 65$^\circ$.}
\begin{tabular}
{rrrrrrrr}
$r$(\AA)&
$E(2\zeta)$&
$E(3\zeta)$&
$E(4\zeta)$&
$E(5\zeta)$&
$E(234)$&
$E(345)$\\ \hline
2.6&642.4&552.6&506.9&---&479.1&\textit{477.1}\\
2.7&454.2&380.5&344.2&---&322.1&\textit{320.6}\\
2.8&318.3&257.4&228.3&---&210.8&\textit{209.6}\\
2.9&220.7&169.9&146.6&137.1&132.5&131.6\\
3.0&150.90&108.37&89.52&81.99&78.25&77.60\\
3.1&101.26&65.47&50.19&44.12&41.09&40.59\\
3.2&66.16&35.92&23.48&18.54&16.12&15.67\\
\end{tabular}
\end{ruledtabular} \end{table} 
\begin{table} \begin{ruledtabular}
\scriptsize \caption{Potential energy $E(r,\theta)$ in meV for the basis sets  $\zeta=$ 2 through 5 and for the  complete-basis-set extrapolations of  Eqs.~(\ref{eq:234})  and  (\ref{eq:345}) for $\theta=$ 70$^\circ$.}
\begin{tabular}
{rrrrrrrr}
$r$(\AA)&
$E(2\zeta)$&
$E(3\zeta)$&
$E(4\zeta)$&
$E(5\zeta)$&
$E(234)$&
$E(345)$\\ \hline
2.6&521.5&434.5&387.1&368.4&358.1&357.6\\
2.7&368.8&297.3&259.4&244.8&236.3&236.2\\
2.8&258.9&199.6&169.3&157.7&150.8&151.0\\
2.9&180.04&130.65&106.21&97.03&91.38&91.69\\
3.0&123.73&82.35&62.56&55.22&50.61&50.95\\
3.1&83.65&48.82&32.74&26.81&23.08&23.35\\
3.2&55.25&25.80&12.69&7.84&4.85&5.02\\
\end{tabular}
\end{ruledtabular} \end{table} 
\begin{table} \begin{ruledtabular}
\scriptsize \caption{Potential energy $E(r,\theta)$ in meV for the basis sets  $\zeta=$ 2 through 5 and for the  complete-basis-set extrapolations of  Eqs.~(\ref{eq:234})  and  (\ref{eq:345}) for $\theta=$ 75$^\circ$.}
\begin{tabular}
{rrrrrrrr}
$r$(\AA)&
$E(2\zeta)$&
$E(3\zeta)$&
$E(4\zeta)$&
$E(5\zeta)$&
$E(234)$&
$E(345)$\\ \hline
2.6&463.6&374.9&323.0&---&291.0&\textit{292.6}\\
2.7&328.6&256.0&214.3&199.2&188.7&190.4\\
2.8&231.7&171.8&138.3&126.4&117.7&119.5\\
2.9&162.35&112.66&85.62&76.17&69.05&70.68\\
3.0&112.90&71.42&49.51&41.95&36.13&37.55\\
3.1&77.70&42.87&25.07&18.96&14.23&15.40\\
3.2&52.69&23.29&8.78&3.77&-0.02&0.86\\
\end{tabular}
\end{ruledtabular} \end{table} 
\begin{table} \begin{ruledtabular}
\scriptsize \caption{Potential energy $E(r,\theta)$ in meV for the basis sets  $\zeta=$ 2 through 5 and for the  complete-basis-set extrapolations of  Eqs.~(\ref{eq:234})  and  (\ref{eq:345}) for $\theta=$ 80$^\circ$.}
\begin{tabular}
{rrrrrrrr}
$r$(\AA)&
$E(2\zeta)$&
$E(3\zeta)$&
$E(4\zeta)$&
$E(5\zeta)$&
$E(234)$&
$E(345)$\\ \hline
2.6&462.2&367.6&308.3&287.1&271.4&274.7\\
2.7&329.1&252.3&204.5&187.9&174.8&178.3\\
2.8&233.8&170.9&132.4&119.4&108.5&111.8\\
2.9&165.68&113.94&82.76&72.49&63.46&66.53\\
3.0&117.17&74.26&48.99&40.80&33.39&36.05\\
3.1&82.63&46.79&26.28&19.67&13.64&15.84\\
3.2&58.02&27.89&11.19&5.79&0.94&2.66\\
\end{tabular}
\end{ruledtabular} \end{table} 
\begin{table} \begin{ruledtabular}
\scriptsize \caption{Potential energy $E(r,\theta)$ in meV for the basis sets  $\zeta=$ 2 through 5 and for the  complete-basis-set extrapolations of  Eqs.~(\ref{eq:234})  and  (\ref{eq:345}) for $\theta=$ 85$^\circ$.}
\begin{tabular}
{rrrrrrrr}
$r$(\AA)&
$E(2\zeta)$&
$E(3\zeta)$&
$E(4\zeta)$&
$E(5\zeta)$&
$E(234)$&
$E(345)$\\ \hline
2.6&516.9&411.7&341.7&---&298.0&\textit{303.2}\\
2.7&370.3&285.8&229.4&---&194.1&\textit{199.2}\\
2.8&265.3&196.9&151.2&136.3&122.7&127.6\\
2.9&190.31&134.54&97.52&85.80&74.41&79.00\\
3.0&136.84&91.00&60.99&51.70&42.28&46.32\\
3.1&98.74&60.74&36.40&28.96&21.25&24.64\\
3.2&71.53&39.77&20.00&13.96&7.72&10.46\\
\end{tabular}
\end{ruledtabular} \end{table} 

\begin{table} \begin{ruledtabular}
\scriptsize \caption{Potential energy $E(r,\theta)$ in meV for the basis sets  $\zeta=$ 2 through 5 and for the  complete-basis-set extrapolations of  Eqs.~(\ref{eq:234})  and  (\ref{eq:345}) for $\theta=$ 90$^\circ$.}
\begin{tabular}
{rrrrrrrr}
$r$(\AA)&
$E(2\zeta)$&
$E(3\zeta)$&
$E(4\zeta)$&
$E(5\zeta)$&
$E(234)$&
$E(345)$\\ \hline
2.6&633.8&512.1&427.7&398.2&374.9&381.1\\
2.7&457.0&360.3&292.3&---&249.6&\textit{255.8}\\
2.8&329.9&252.6&197.5&179.5&162.9&169.1\\
2.9&238.9&176.6&131.8&---&103.7&\textit{109.6}\\
3.0&173.85&123.19&86.90&75.93&64.10&69.58\\
3.1&127.45&85.87&56.42&---&37.95&\textit{42.66}\\
3.2&94.26&59.79&35.91&28.93&20.94&24.89\\
\end{tabular}
\end{ruledtabular} \end{table} 
\begin{table} \begin{ruledtabular}
\scriptsize \caption{Potential energy $E(r,\theta)$ in meV for the basis sets  $\zeta=$ 2 through 5 and for the  complete-basis-set extrapolations of  Eqs.~(\ref{eq:234})  and  (\ref{eq:345}) for $\theta=$ 95$^\circ$.}
\begin{tabular}
{rrrrrrrr}
$r$(\AA)&
$E(2\zeta)$&
$E(3\zeta)$&
$E(4\zeta)$&
$E(5\zeta)$&
$E(234)$&
$E(345)$\\ \hline
2.6&826.0&679.8&576.3&---&511.4&\textit{517.9}\\
2.7&598.8&484.1&400.9&---&348.5&\textit{355.6}\\
2.8&434.6&344.0&276.7&---&234.3&\textit{241.6}\\
2.9&316.6&244.4&189.8&---&155.4&\textit{162.6}\\
3.0&232.0&174.0&129.7&116.2&101.7&108.4\\
3.1&171.46&124.46&88.42&77.91&65.68&71.82\\
3.2&128.13&89.57&60.35&52.02&41.92&47.20\\
\end{tabular}
\end{ruledtabular} \end{table} 
\begin{table} \begin{ruledtabular}
\scriptsize \caption{Potential energy $E(r,\theta)$ in meV for the basis sets  $\zeta=$ 2 through 5 and for the  complete-basis-set extrapolations of  Eqs.~(\ref{eq:234})  and  (\ref{eq:345}) for $\theta=$ 100$^\circ$.}
\begin{tabular}
{rrrrrrrr}
$r$(\AA)&
$E(2\zeta)$&
$E(3\zeta)$&
$E(4\zeta)$&
$E(5\zeta)$&
$E(234)$&
$E(345)$\\ \hline
2.6&1\,114.1&932.8&804.2&---&723.5&\textit{730.8}\\
2.7&811.2&670.6&567.5&---&502.6&\textit{510.6}\\
2.8&591.0&481.2&398.1&---&345.7&\textit{354.0}\\
2.9&431.7&345.4&278.2&---&235.7&\textit{243.9}\\
3.0&317.2&248.7&194.1&---&159.5&\textit{167.4}\\
3.1&235.0&180.1&135.8&122.6&107.7&115.0\\
3.2&176.09&131.60&95.62&85.34&72.81&79.40\\
\end{tabular}
\end{ruledtabular} \end{table} 
\begin{table} \begin{ruledtabular}
\scriptsize \caption{Potential energy $E(r,\theta)$ in meV for the basis sets  $\zeta=$ 2 through 5 and for the  complete-basis-set extrapolations of  Eqs.~(\ref{eq:234})  and  (\ref{eq:345}) for $\theta=$ 105$^\circ$.}
\begin{tabular}
{rrrrrrrr}
$r$(\AA)&
$E(2\zeta)$&
$E(3\zeta)$&
$E(4\zeta)$&
$E(5\zeta)$&
$E(234)$&
$E(345)$\\ \hline
2.6&1\,528.2&1\,297.6&1\,135.9&---&1\,034.5&\textit{1\,045.2}\\
2.7&1\,116.8&939.6&810.5&---&729.4&\textit{739.0}\\
2.8&815.5&678.6&575.1&539.5&509.9&518.8\\
2.9&596.5&490.1&406.6&378.7&353.9&362.5\\
3.0&438.1&354.7&287.2&265.5&244.5&252.9\\
3.1&324.2&258.2&203.4&186.6&168.7&176.9\\
3.2&242.4&189.6&145.1&132.1&116.9&124.6\\
\end{tabular}
\end{ruledtabular} \end{table} 
\begin{table} \begin{ruledtabular}
\scriptsize \caption{Potential energy $E(r,\theta)$ in meV for the basis sets  $\zeta=$ 2 through 5 and for the  complete-basis-set extrapolations of  Eqs.~(\ref{eq:234})  and  (\ref{eq:345}) for $\theta=$ 112.5$^\circ$.}
\begin{tabular}
{rrrrrrrr}
$r$(\AA)&
$E(2\zeta)$&
$E(3\zeta)$&
$E(4\zeta)$&
$E(5\zeta)$&
$E(234)$&
$E(345)$\\ \hline
2.6&2\,479.0&2\,136.8&1\,905.7&---&1\,761.3&\textit{1\,772.8}\\
2.7&1\,820.2&1\,559.6&1\,376.0&---&1\,260.9&\textit{1\,272.6}\\
2.8&1\,332.6&1\,133.9&987.7&---&895.6&\textit{907.4}\\
2.9&974.8&822.6&705.8&---&632.0&\textit{643.5}\\
3.0&714.3&597.1&503.2&---&443.8&\textit{454.8}\\
3.1&526.0&434.7&359.0&---&311.0&\textit{321.3}\\
3.2&390.4&318.6&257.3&---&218.4&\textit{227.9}\\
\end{tabular}
\end{ruledtabular} \end{table} 
\begin{table} \begin{ruledtabular}
\scriptsize \caption{Potential energy $E(r,\theta)$ in meV for the basis sets  $\zeta=$ 2 through 5 and for the  complete-basis-set extrapolations of  Eqs.~(\ref{eq:234})  and  (\ref{eq:345}) for $\theta=$ 120$^\circ$.}
\begin{tabular}
{rrrrrrrr}
$r$(\AA)&
$E(2\zeta)$&
$E(3\zeta)$&
$E(4\zeta)$&
$E(5\zeta)$&
$E(234)$&
$E(345)$\\ \hline
2.6&4\,000.9&3\,484.8&3\,152.6&---&2\,945.8&\textit{2\,967.1}\\
2.7&2\,950.9&2\,558.8&2\,295.4&---&2\,130.8&\textit{2\,148.6}\\
2.8&2\,166.1&1\,869.0&1\,660.2&---&1\,529.3&\textit{1\,544.3}\\
2.9&1\,585.1&1\,359.8&1\,194.1&---&1\,089.8&\textit{1\,102.7}\\
3.0&1\,158.8&987.3&855.4&809.9&772.1&783.4\\
3.1&848.6&717.1&611.6&---&544.8&\textit{555.1}\\
3.2&624.3&522.6&437.8&409.6&384.0&393.2\\
\end{tabular}
\end{ruledtabular} \end{table} 
\begin{table} \begin{ruledtabular}
\scriptsize \caption{Potential energy $E(r,\theta)$ in meV for the basis sets  $\zeta=$ 2 through 5 and for the  complete-basis-set extrapolations of  Eqs.~(\ref{eq:234})  and  (\ref{eq:345}) for $\theta=$ 135$^\circ$.}
\begin{tabular}
{rrrrrrrr}
$r$(\AA)&
$E(2\zeta)$&
$E(3\zeta)$&
$E(4\zeta)$&
$E(5\zeta)$&
$E(234)$&
$E(345)$\\ \hline
2.6&9\,839.6&8\,706.9&8\,032.1&---&7\,614.9&\textit{7\,668.2}\\
2.7&7\,316.4&6\,446.8&5\,915.6&---&5\,586.5&\textit{5\,629.5}\\
2.8&5\,405.9&4\,743.7&4\,325.3&---&4\,065.2&\textit{4\,099.7}\\
2.9&3\,971.2&3\,470.0&3\,140.5&---&2\,935.0&\textit{2\,962.3}\\
3.0&2\,903.1&2\,525.3&2\,265.8&---&2\,103.3&\textit{2\,124.9}\\
3.1&2\,115.0&1\,830.3&1\,625.8&1\,555.3&1\,497.3&1\,514.4\\
3.2&1\,538.7&1\,323.4&1\,161.8&---&1\,059.9&\textit{1\,073.8}\\
\end{tabular}
\end{ruledtabular} \end{table} 
\begin{table} \begin{ruledtabular}
\scriptsize \caption{Potential energy $E(r,\theta)$ in meV for the basis sets  $\zeta=$ 2 through 5 and for the  complete-basis-set extrapolations of  Eqs.~(\ref{eq:234})  and  (\ref{eq:345}) for $\theta=$ 150$^\circ$.}
\begin{tabular}
{rrrrrrrr}
$r$(\AA)&
$E(2\zeta)$&
$E(3\zeta)$&
$E(4\zeta)$&
$E(5\zeta)$&
$E(234)$&
$E(345)$\\ \hline
2.6&21\,162.&18\,974.&17\,686.&---&16\,890.&\textit{16\,958.}\\
2.7&15\,781.&14\,088.&13\,088.&---&12\,470.&\textit{12\,526.}\\
2.8&11\,709.3&10\,408.6&9\,632.6&---&9\,152.8&\textit{9\,199.0}\\
2.9&8\,641.0&7\,649.4&7\,045.7&---&6\,671.8&\textit{6\,709.5}\\
3.0&6\,341.5&5\,591.0&5\,120.1&---&4\,827.7&\textit{4\,858.2}\\
3.1&4\,629.6&4\,064.9&3\,697.1&---&3\,467.9&\textit{3\,492.5}\\
3.2&3\,364.8&2\,941.3&2\,653.7&---&2\,473.8&\textit{2\,493.5}\\
\end{tabular}
\end{ruledtabular} \end{table} 
\begin{table} \begin{ruledtabular}
\scriptsize \caption{Potential energy $E(r,\theta)$ in meV for the basis sets  $\zeta=$ 2 through 5 and for the  complete-basis-set extrapolations of  Eqs.~(\ref{eq:234})  and  (\ref{eq:345}) for $\theta=$ 165$^\circ$.}
\begin{tabular}
{rrrrrrrr}
$r$(\AA)&
$E(2\zeta)$&
$E(3\zeta)$&
$E(4\zeta)$&
$E(5\zeta)$&
$E(234)$&
$E(345)$\\ \hline
2.6&36\,322.&32\,819.&30\,798.&---&29\,553.&\textit{29\,555.}\\
2.7&27\,068.&24\,356.&22\,773.&---&21\,795.&\textit{21\,798.}\\
2.8&20\,075.&17\,996.&16\,775.&---&16\,022.&\textit{16\,026.}\\
2.9&14\,823.&13\,241.&12\,304.&---&11\,725.&\textit{11\,730.}\\
3.0&10\,894.5&9\,698.9&8\,977.9&---&8\,531.7&\textit{8\,537.0}\\
3.1&7\,968.5&7\,069.6&6\,513.0&---&6\,167.7&\textit{6\,173.4}\\
3.2&5\,800.9&5\,128.0&4\,697.1&---&4\,428.9&\textit{4\,434.8}\\
\end{tabular}
\end{ruledtabular} \end{table} 
\begin{table} \begin{ruledtabular}
\scriptsize \caption{Potential energy $E(r,\theta)$ in meV for the basis sets  $\zeta=$ 2 through 5 and for the  complete-basis-set extrapolations of  Eqs.~(\ref{eq:234})  and  (\ref{eq:345}) for $\theta=$ 175$^\circ$.}
\begin{tabular}
{rrrrrrrr}
$r$(\AA)&
$E(2\zeta)$&
$E(3\zeta)$&
$E(4\zeta)$&
$E(5\zeta)$&
$E(234)$&
$E(345)$\\ \hline
2.6&43\,155.&39\,081.&36\,804.&---&35\,406.&\textit{35\,497.}\\
2.7&32\,313.&29\,128.&27\,289.&---&26\,156.&\textit{26\,232.}\\
2.8&23\,961.&21\,525.&20\,106.&---&19\,230.&\textit{19\,291.}\\
2.9&17\,676.&15\,830.&14\,745.&---&14\,075.&\textit{14\,124.}\\
3.0&12\,983.&11\,593.&10\,763.&---&10\,251.&\textit{10\,289.}\\
3.1&9\,494.2&8\,451.7&7\,815.5&---&7\,421.3&\textit{7\,451.2}\\
3.2&6\,912.3&6\,133.3&5\,643.3&---&5\,338.9&\textit{5\,361.5}\\
\end{tabular}
\end{ruledtabular} \end{table} 
\begin{table} \begin{ruledtabular}
\scriptsize \caption{Potential energy $E(r,\theta)$ in meV for the basis sets  $\zeta=$ 2 through 5 and for the  complete-basis-set extrapolations of  Eqs.~(\ref{eq:234})  and  (\ref{eq:345}) for $\theta=$ 180$^\circ$.} \label{table:sup180}
\begin{tabular}
{rrrrrrrr}
$r$(\AA)&
$E(2\zeta)$&
$E(3\zeta)$&
$E(4\zeta)$&
$E(5\zeta)$&
$E(234)$&
$E(345)$\\ \hline
2.6&44\,099.&39\,951.&37\,646.&---&36\,230.&\textit{36\,333.}\\
2.7&33\,056.&29\,803.&27\,930.&---&26\,776.&\textit{26\,860.}\\
2.8&24\,516.&22\,029.&20\,581.&---&19\,688.&\textit{19\,757.}\\
2.9&18\,082.&16\,198.&15\,093.&---&14\,410.&\textit{14\,465.}\\
3.0&13\,279.&11\,862.&11\,017.&---&10\,495.&\textit{10\,538.}\\
3.1&9\,710.0&8\,647.5&8\,000.3&---&7\,599.4&\textit{7\,632.4}\\
3.2&7\,069.3&6\,275.5&5\,777.5&---&5\,468.1&\textit{5\,492.9}\\
\end{tabular}
\end{ruledtabular} \end{table}

\begin{table*}
\begin{ruledtabular}
\scriptsize
\caption{$\zeta=5$ 
contraction coefficients for the 
twelve
S 
states} 
\label{table:S5Z}
\begin{tabular}
{rrrrrrrrrrrrr}
$\zeta$&1&2&3&4&5&6&7&8&9&10&11&12\\
\hline
7.374E7&-6.900E-5&3.100E-5&1.400E-5&-6.000E-6&-2.000E-6&-2.000E-6&1.950E-4&0&0&0&0&0\\
-1.948E7&-7.200E-5&3.300E-5&1.500E-5&-6.000E-6&-2.000E-6&-2.000E-6&2.050E-4&0&0&0&0&0\\
-6.560E6&-2.040E-4&9.200E-5&4.300E-5&-1.700E-5&-5.000E-6&-5.000E-6&5.800E-4&0&0&0&0&0\\
-2.431E6&-3.180E-4&1.440E-4&6.700E-5&-2.600E-5&-7.000E-6&-7.000E-6&9.020E-4&0&0&0&0&0\\
-9.701E5&-6.440E-4&2.920E-4&1.360E-4&-5.300E-5&-1.500E-5&-1.500E-5&1.825E-3&0&0&0&0&0\\
-4.055E5&-1.111E-3&5.030E-4&2.340E-4&-9.200E-5&-2.600E-5&-2.600E-5&3.140E-3&0&0&0&0&0\\
-1.766E5&-2.058E-3&9.330E-4&4.350E-4&-1.710E-4&-4.800E-5&-4.800E-5&5.790E-3&0&0&0&0&0\\
\end{tabular}
\end{ruledtabular}
\end{table*}

\begin{table*}
\begin{ruledtabular}
\scriptsize
\caption{$\zeta=5$ 
contraction coefficients for the 
ten
P 
states} 
\label{table:P5Z}
\begin{tabular}
{rrrrrrrrrrr}
$\zeta$&1&2&3&4&5&6&7&8&9&10\\
\hline
-3.764E7&-2.000E-6&1.000E-6&0.000E0&3.000E-6&0.000E0&0&0&0&0&0\\
-8.288E6&-3.000E-6&1.000E-6&1.000E-6&7.000E-6&0.000E0&0&0&0&0&0\\
-2.187E6&-9.000E-6&4.000E-6&1.000E-6&1.900E-5&0.000E0&0&0&0&0&0\\
-6.494E5&-2.200E-5&1.000E-5&4.000E-6&4.600E-5&-1.000E-6&0&0&0&0&0\\
-2.121E5&-5.500E-5&2.500E-5&9.000E-6&1.150E-4&-2.000E-6&0&0&0&0&0\\
-7.507E4&-1.380E-4&6.300E-5&2.200E-5&2.890E-4&-5.000E-6&0&0&0&0&0\\
-2.852E4&-3.620E-4&1.650E-4&5.900E-5&7.560E-4&-1.400E-5&0&0&0&0&0\\
\end{tabular}
\end{ruledtabular}
\end{table*}

\begin{table}
\begin{ruledtabular}
\scriptsize
\caption{$\zeta=5$ 
contraction coefficients for the 
eight
D 
states} 
\label{table:D5Z}
\begin{tabular}
{rrrrrrrrr}
$\zeta$&1&2&3&4&5&6&7&8\\
\hline
-3.415E4&7.000E-6&-3.000E-6&1.000E-6&0&0&0&0&0\\
-8.451E3&5.200E-5&-2.300E-5&4.000E-6&0&0&0&0&0\\
-2.922E3&2.900E-4&-1.280E-4&2.400E-5&0&0&0&0&0\\
-1.205E3&1.366E-3&-6.070E-4&1.130E-4&0&0&0&0&0\\
-5.543E2&5.549E-3&-2.472E-3&4.600E-4&0&0&0&0&0\\
-2.752E2&1.899E-2&-8.527E-3&1.585E-3&0&0&0&0&0\\
-1.439E2&5.376E-2&-2.433E-2&4.531E-3&0&0&0&0&0\\
-7.815E1&1.246E-1&-5.738E-2&1.070E-2&0&0&0&0&0\\
-4.358E1&2.256E-1&-1.036E-1&1.932E-2&0&0&0&0&0\\
-2.463E1&3.084E-1&-1.371E-1&2.546E-2&0&0&0&0&0\\
-1.410E1&2.892E-1&-1.017E-1&1.805E-2&0&0&0&0&0\\
-8.084E0&1.572E-1&7.419E-2&-1.885E-2&0&0&0&0&0\\
-4.582E0&4.125E-2&2.999E-1&-6.481E-2&0&0&0&0&0\\
-2.565E0&3.617E-3&3.988E-1&-8.785E-2&0&0&0&0&0\\
-1.412E0&1.610E-4&2.811E-1&-4.917E-2&0&0&0&0&0\\
-7.452E-1&-7.800E-5&9.745E-2&7.712E-2&0&0&0&0&0\\
-3.523E-1&-4.000E-6&1.073E-2&2.480E-1&1&0&0&0&0\\
-1.508E-1&-6.000E-6&-5.000E-5&3.827E-1&0&1&0&0&0\\
-6.184E-2&2.000E-6&9.200E-5&3.962E-1&0&0&1&0&0\\
-2.463E-2&-1.000E-6&-2.900E-5&1.741E-1&0&0&0&1&0\\
-9.850E-3&0&0&0&0&0&0&0&1\\
\end{tabular}
\end{ruledtabular}
\end{table}

\begin{table}
\begin{ruledtabular}
\scriptsize
\caption{$\zeta=5$ 
contraction coefficients for the 
five
F 
states} 
\label{table:F5Z}
\begin{tabular}
{rrrrrr}
$\zeta$&1&2&3&4&5\\
\hline
-1.132E1&3.644E-2&0&0&0&0\\
-3.399E0&5.892E-2&0&0&0&0\\
-9.299E-1&4.237E-1&0&0&0&0\\
-4.902E-1&5.299E-1&1&0&0&0\\
-2.200E-1&1.207E-1&0&1&0&0\\
-7.491E-2&6.011E-3&0&0&1&0\\
-2.996E-2&0&0&0&0&1\\
\end{tabular}
\end{ruledtabular}
\end{table}

\begin{table}
\begin{ruledtabular}
\scriptsize
\caption{$\zeta=5$ 
contraction coefficients for the 
three 
G 
states} 
\label{table:G5Z}
\begin{tabular}
{rrrr}
$\zeta$&1&2&3\\
\hline
-3.497E-1&1&0&0\\
-9.820E-2&0&1&0\\
-3.930E-2&0&0&1\\
\end{tabular}
\end{ruledtabular}
\end{table}

\begin{table}
\begin{ruledtabular}
\scriptsize
\caption{$\zeta=5$
contraction coefficients for the
two 
H
states} 
\label{table:H5Z}
\begin{tabular}
{rrr}
$\zeta$&1&2\\
\hline
-1.287E-1&1&0\\
-5.150E-2&0&1\\
\end{tabular}
\end{ruledtabular}
\end{table}

\end{document}